\documentclass[twocolumn]{article}
\usepackage{amsmath, amssymb}
\usepackage[utf8]{inputenc} 
\usepackage[T1]{fontenc}    
\usepackage{graphicx}
\usepackage{hyperref}
\usepackage[numbers]{natbib}
\usepackage{comment}
\usepackage{appendix}
\usepackage{booktabs}
\usepackage{float}
\usepackage{bm}
\usepackage{xcolor}
\usepackage{hyperref}     
\usepackage{doi}          
\hypersetup{colorlinks=true, linkcolor=blue, citecolor=cyan}

\title{Clifford Accelerated Adaptive QAOA}
\author{Théo Lisart-Liebermann, Arcesio Castanadena Medina \\
        Fraunhofer ITWM \\
        \texttt{theo.lisart@itwm.fraunhofer.de}}

\begin{document}

\maketitle

\begin{abstract}
Clifford Circuit Initializaton improves on initial guess of parameters on Parametric Quantum Circuits (PQCs) by leveraging efficient simulation of circuits made out of gates from the Clifford Group. The parameter space is pre-optimized by exploring the Hilbert space in a reduced ensemble of Clifford-expressible points (Clifford Points), providing better initialization. Simultaneously, dynamical circuit reconfiguration algorithms, such as ADAPT-QAOA, improve on QAOA performances by providing a gate re-configuration routine while the optimization is being executed. In this article, we show that Clifford Point approximations at multiple levels of ADAPT allow for multiple improvements while increasing quantum-classical integration opportunities. First we show numerically that Clifford Point preoptimization offers non-trivial gate-selection behavior in ADAPT with some possible convergence improvement. Second, that Clifford Point approximations allows for more suited, fully parallel and fully classical ADAPT operator selection for MaxCut and the TFIM problem. Finally, we show that applying 10 to 30\% error approximation on T-gates using low-rank stabilizer decomposition can provide significative improvements in convergence quality for the MaxCut and TFIM problem. The latter hints at significant T-gate over-representation in antsatz design, opening opportunities for aggressive compilation optimizations.
\end{abstract}

\section{Introduction}

Providing an improved starting point for Parametric Quantum Circuits (PQCs) optimization is an active area of research. Proposed approaches include classical pre-evaluation of the solution landscape~\cite{Egger_2021} or efficient resolution of a relaxed version of the problem~\cite{sridhar2023adaptqaoaclassicallyinspiredinitial}. In the case of the MaxCut problem, pre-optimized QAOA reportedly outperforms the state-of-the-art Goemans-Williamson algorithm at low circuit depth~\cite{Tate_2023}. However, those methods rely on prior knowledge of the solution structure, which limits their general applicability. In contrast, discrete methods searching the Stabilizer sub-space of the parametrized antsatz, using Bayesian optimization~\cite{ravi2023cafqaclassicalsimulationbootstrap} or simulated annealing~\cite{chengphysreview}, do not require any insight into the solution.

One of the strictest conditions of success for QAOA lies in the general trainability-expressivity trade-off of PQCs design: overparametrization introduces objective function non-convexities, while underparametrization reduces the coverage of accessible solutions~\cite{funcke2021dimensionalexpressivityanalysisbestapproximation, Gago_Encinas_2023}. Methods like ADAPT-QAOA, improve on it by adapting the topology of the mixer layer on the QAOA antsatz as the classical optimization routine is executed~\cite{zhu2022adaptivequantumapproximateoptimization, Yanakiev24}.

The Gottesmann-Knill theorem states that if any quantum process performs only (i) Clifford group gates, (ii) measurements of Pauly group operators and (iii) Clifford group operations conditioned on classical bits, then the results of earlier measurements can be simulated in polynomial time on a probabilistic classical computer~\cite{gottesman1998heisenbergrepresentationquantumcomputers}. This theorem indicates that a subgroup of a QAOA antsatz can be identified as a Clifford-only circuit, or approximately Clifford, and be efficiently simulated on classical hardware.

This work extends the concepts of~\cite{chengphysreview} and tools introduced in~\cite{Bravyi_2016} to ADAPT-QAOA. The introduction of Clifford approximations at multiple levels in ADAPT allows for further quantum-classical integration into existing computer infrastructure, aiming at reducing QPU calls when they are not improving on computation speed or cost. Following this approach, we explore three potential candidates : Clifford Point preoptimization using discrete methods, Clifford Point approximation in the operator selection routine within ADAPT, and finally, the use of low-rank stabilizer decomposition for circuit evaluation.

Our article is structured as the following: first we explore in Sec.~\ref{sec:theoretical_background} the required theoretical background for ADAPT-QAOA solving MaxCut and ground state discovery of the Transverse Field Ising Model (TFIM); we then expose the two approximations we apply on PQCs as reduction to Clifford-only circuits, namely the Clifford Point approximation and the low-rank stabilizer decomposition. We then in Sec.~\ref{sec:results} present numerical results for their use in practice. Finally in Sec.~\ref{sec:conclusion} we conclude and propose some ideas for further research.

\section{Theoretical considerations}
\label{sec:theoretical_background}
In the following section, we introduce the required concepts needed to present our results. First, an introduction to the the ADAPT-QAOA procedure for MaxCut and the TFIM, followed by a short introduction to efficient Clifford circuit simulation. Finally, we cover the Clifford Point approximation and stabilizer low-rank decomposition.

\begin{figure*}
    \centering
    \vspace{-2.5cm}
    \label{fig:base_explanation_adapt}
    \includegraphics[scale=0.85]{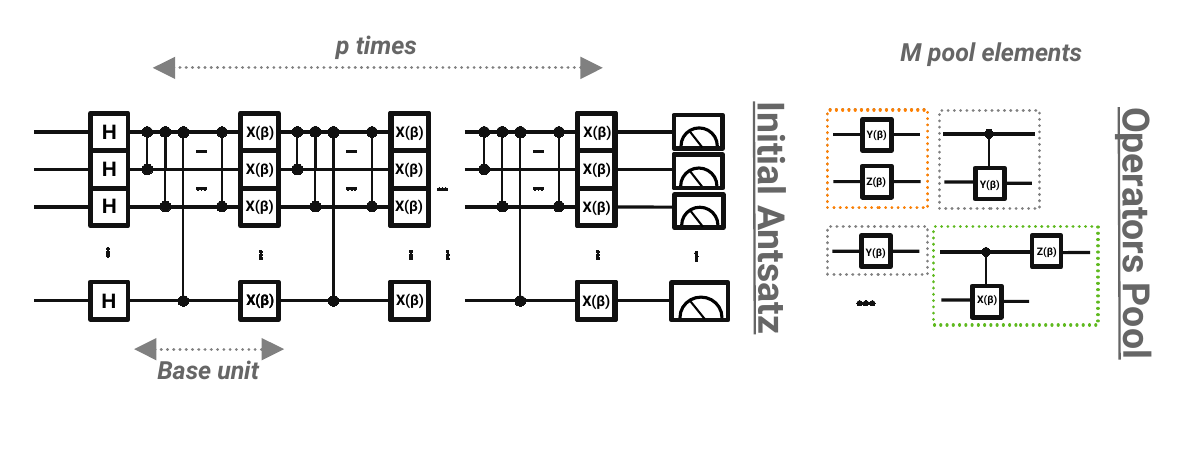}
    \caption{Example base antsatz, where the RZZ gates are one-to-one representation of the graph in the Z basis. Additionally, the operator pool is defined through SAT considerations. For simplicity, we wrote any gate rotation as $A(\beta)$, where $A\in[X, Y, Z]$, and $\beta=(\beta_1, ...\beta_n)$ where each parametric gate $A$ is associated an independent $\beta_i$ }
\end{figure*}

\subsection{ADAPT-QAOA}
ADAPT was first introduced in the context of VQE for molecular simulations~\cite{Grimsley_2019}, then modified for the QAOA procedure~\cite{zhu2022adaptivequantumapproximateoptimization}. In principle, the method starts from a standard QAOA Hamiltonian~\cite{farhi2014quantumapproximateoptimizationalgorithm}, then brings sequential improvements by adding parametrized gates iteratively to the mixer layer.

Departing from the original QAOA~\cite{farhi2014quantumapproximateoptimizationalgorithm}, the mixer layer in the throtterized adiabatic evolution is formally replaced by a set of mixers defined layer by layers~\cite{zhu2022adaptivequantumapproximateoptimization}:

\begin{equation}
    \label{eq:throtterizer}
    |\psi_p(\gamma, \beta)\rangle = \big(\prod_{k=1}^p e^{-i\hat{A}_j\beta_k}e^{-i\hat{H}_C\gamma_k}\big)|\psi_{ref}\rangle\,,
\end{equation}

Where $p$ is the throtter precision, or repeat of the alternating operator, the couple $(\beta_k, \gamma_k)$ the parameter vectors, and $\hat{A}_j$ an operator picked from the pool $\{\hat{A}_j\}$ at the $j^{th}$ iteration of ADAPT. The entire ADAPT procedure can be then summarized as follow, simplified as a flow diagram at~Fig.~\ref{fig:flow_diagram_adapt_full}:

\begin{enumerate}
        \item Define the operator set $\{A_j\}$, the mixer-pool, and select a suitable reference state, $|\psi_{ref}\rangle = |+ \rangle^{\otimes n}$
        \item Prepare $|\psi^{k-1}\rangle$ on the QPU, and measure the energy gradient with respect to the pool (1) with
        \begin{equation}
        \label{eq:decision_problem_eval}
                -i\langle\psi^{k-1}|e^{i\hat{H}_C\gamma_k}[\hat{H}_C,\hat{A}_j]e^{-i\hat{H}_C\gamma_k}|\psi^{k-1}\rangle
        \end{equation}
        This means that for each layer of the mixer, we compute the energy gradient and associate it in a $i=1,...j$ vector. And the new variational parameter $\gamma_k$ is set to a predefined value $\gamma_0$

        \item From Eq~\ref{eq:decision_problem_eval} be build a vector $E^k$, and from it pick an operator from $\{\hat{A}_j\}$. We modify the antsatz by adding $\hat{A}_{max}^k$ associated with the largest component of the gradient $|\psi^k\rangle = e^{-i\hat{A}_{max}^k\beta_k}e^{-i\hat{H}_C\gamma_k}|\psi^{k-1}\rangle$. A stopping condition can be introduced: $grad(E^k)$ is below a preset threshold, ADAPT is stopped as adding further parametric gates will not improve expressivity. If the threshold is not met:
        \item Optimize all parameters $\beta_m, \gamma_m$ such as $\langle\psi^k|\hat{H}_C|\psi^k\rangle$ is minimized, then go back to (2).
\end{enumerate}

In the original article~\cite{zhu2022adaptivequantumapproximateoptimization}, the operator pool as well as the optimal choice for the newly added parameter $\gamma_0$ are provided through physical considerations around the concept of Shortcut to Adiabaticity (STA)~\cite{Gu_ry_Odelin_2019}. For our purpose, this means $\gamma_0\xrightarrow[]{}0$ and the operator pool:

\begin{equation}
\small
\label{eq:operator_pool}
\left\{
\begin{aligned}
    \hat{P}_{QAOA} &= \big\{\sum_{i\in Q}\hat{X}_i\big\} \\
    \hat{P}_{single} &= \bigcup_{i\in Q}\big\{\hat{X}_i, \hat{Y}_i\big\} \cup \big\{\sum_{i\in Q}\hat{Y}_i\big\}\cup \hat{P}_{QAOA} \\
    \hat{P}_{multi} &= \bigcup_{i,j\in Q\times Q}\big\{\hat{B}_i\hat{C}_j\,|\,\hat{B}_i, \hat{C}_j\in \big\{\hat{X}, \hat{Y}, \hat{Z}\big\}\big\}\cup \hat{P}_{single}
\end{aligned}
\right.
\end{equation}
We solve optimization problems of interest by mapping them to the Hamiltonian $\hat{H}_C$, which is the subject of the next section.

\subsection{MaxCut and TFIM}
We implement the MaxCut problem for any connected graphs, the Transverse Field Ising Model (TFIM) ground state problem on the other second.

\paragraph{MaxCut:} Given a weighted graph \( G = (V, E, w) \) where \( V \) is the set of vertices, \( E \) is the set of edges, and \( w: E \rightarrow \mathbb{R}^+ \) is a weight function assigning a positive weight \( w_{ij} \) to each edge \( (i, j) \in E \), the MaxCut problem is fiding a partition of the vertex set \( V \) into two disjoint subsets \( S \) and \( \bar{S} = V \setminus S \) such that the sum of the weights of the edges between the two subsets is maximized. Assigning a binary variable \( z_i \in \{0, 1\} \) for each vertex \( i \in V \), the formal binary expression of the objective function the problem is:

\begin{equation}
\label{eq:classical_cost_function}
\text{MaxCut}(G) = \max_{\mathbf{z} \in \{0, 1\}^n} \sum_{(i,j) \in E} w_{ij} \cdot \frac{1}{2} (1 - z_i z_j)\,,
\end{equation}

where, \( z_i = 0 \) or \( z_i = 1 \) represents the assignment of vertex \( i \) to one of the two subsets, and \( n = |V| \) is the number of vertices in the graph. For a given bitstring, the weighted loss function for a $x$ length $n$ bitstring: $C(x) = \sum_{(i,j) \in E} w_{ij} \cdot \frac{1}{2} (x_i x_j - 1)$, where we express (\ref{eq:classical_cost_function}) as a minimization problem.

To solve this problem using a quantum-classical approach, the QAOA method uses a $n$ qubit antsatz trotterizing the adiabatic evolution~\cite{farhi2000quantumcomputationadiabaticevolution}. The precision of the throtterizer is determined by the parameter $p>0$ that defines the number of repeating units of Mixer and Problem layers in Eq.~\ref{eq:throtterizer}. We can trivially map binary variables $\{z_i\}^{0, 1}$ to eigenvalues of $\{\hat{Z}_i\}^{-1, 1}$, translating directly the cost function (\ref{eq:classical_cost_function}) to Pauli-Z operators acting on each qubit:

\begin{equation}
\label{eq:quantum_cost_function}
\hat{P} = \sum_{(i,j) \in E} w_{ij} \cdot \frac{1}{2} (\hat{Z}_i \hat{Z}_j - \hat{I})\,.
\end{equation}

This expression provides a formal definition of both mapping to the problem layer to parametric rotation-controlled-Z gates, as well as an objective function from measured distributions. Interestingly, this later expression is a special case of a more general quantum control problem, the so-called Transverse Field Ising Model.

\paragraph{TFIM:} is an ubiquitous model appearing in quantum phase transition~\cite{Heyl_2013, PhysRevB.96.134427} and condensed matter physics~\cite{PhysRev.127.1508, PFEUTY197079}. It is the quantum equivalent to Ising chains in statistical mechanics, and has been thoroughly studied due to its simple one dimensional description. It is mainly used to describe simple spin systems along the chosen $Z$ direction, implementing an interaction with a transverse field competing with the interaction terms, potentially a longitudinal field along the $X$ axis. The Hamiltonian has the form:

\begin{equation}
    \label{eq:general_tfim_hamiltonian}
    \hat{H} = \sum_{i, j} w_{ij}\hat{Z}_i\hat{Z}_j + \sum_{i} g_x\hat{X}_i + g_z\hat{Z}_i\,,
\end{equation}

where we introduce the external coupling interaction couple $(g_x, g_z)$ respectively the longitudinal and transverse coupling strength, $w_{ij}$ matrix elements of the interaction matrix $W$.

\paragraph{Solving with QAOA:} one can solve both problems using QAOA by defining the problem Hamiltonian in the optimization problem $min_{\beta_m, \gamma_m}\langle\psi^k|\hat{H}_C|\psi^k\rangle$, and mapping single spin operators $\hat{X}_i$ and $\hat{Z}_i$ to parametrized rotation gates on the $i^{th}$ qubits on a quantum processor. In addition, one can map $\hat{Z}_i\hat{Z}_j$ to a parametrized controlled-Z gate (RRZ) gate with control qubit $i$ and target qubit $j$.

What is understood as continuous optimization will be a standard parameters optimization using SPSA~\cite{spall92}. Ground states and exact solutions are found for MaxCut by brute-forcing up to 6 nodes, or a simple greedy algorithm at higher dimensions. The ground state of the TFIM is found by direct diagonalization of the Hamiltonian using traditional pivot method.

\subsection{Efficient classical simulation of quantum circuits}
Quantum circuits can be efficiently simulated on classical hardware if constituted from a set of gates coming from the Clifford group, to the limit that all measurements are in the Pauli basis~\cite{gottesman1998heisenbergrepresentationquantumcomputers}. A general phase-rotation gate can be written~\cite{Nielsen_Chuang_2010}:

\begin{equation}
\label{eq:general_rotation}
    \hat{R}(\vec{n}, \theta) = \exp\left(-i\frac{\theta}{2} \, \vec{n} \cdot \vec{\hat{\sigma}}\right)\,,
\end{equation}

where in Eq.~\eqref{eq:general_rotation}, $\vec{n}$ is a unit vector specifying the axis of rotation, $\theta$ is the rotation angle, and $\vec{\hat{\sigma}} = (\hat{X}, \hat{Y}, \hat{Z})$ denotes the vector of Pauli operators.

Libraries such as Stim~\cite{Grimsley_2019} have been developed for high speed analysis of purely Clifford circuit. Most of those methods rely on Clifford Tableau methods, which is also implemented in \textit{qiskit} through \textit{stabilizer} simulator. Modern techniques allow for high sampling rates of Stabilizer circuits~\cite{PhysRevA.70.052328}, which is an advantage if part of the quantum procedure can be executed on classical hardware.

\subsection{Clifford approximation of quantum circuits}
\begin{figure*}[t]
    \centering
    \vspace{-4cm}
    \includegraphics[scale=0.65]{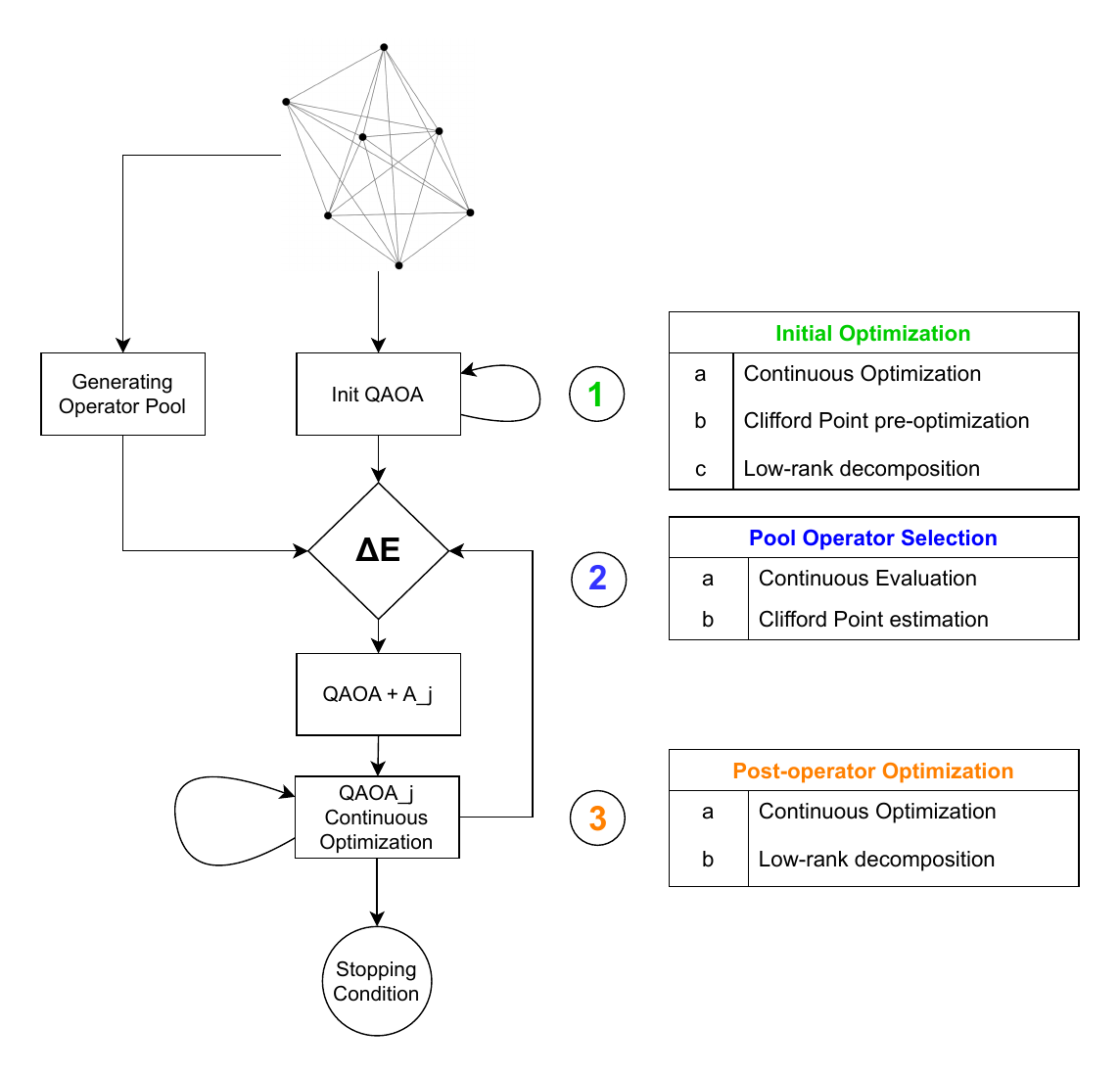}
    \caption{Block flow diagram of the ADAPT-QAOA procedure with initial mapping of a graph for MaxCut evaluation. TFIM problems are solved using the same procedure. Each interest points is highlighted, where approximations can be introduced.}
    \label{fig:flow_diagram_adapt_full}
\end{figure*}

The following approach relies on introducing Clifford circuit approximations into various points in the ADAPT procedure in Fig.~(\ref{fig:flow_diagram_adapt_full}). Within this work, we focus on Clifford Point approximations~\cite{chengphysreview} and low-rank stabilizer decomposition~\cite{Bravyi_2016, Bravyi_2019}. The first one relies on the exploration of the reduced Hilbert space through discrete exploration of the Clifford Points in the parameter space. The second approximation relies on Clifford+T~\cite{Bravyi_2019} so-called low-rank stabilizer decomposition, allowing for better scaling of quantum circuit simulation by decomposing the full circuit into a controllable approximation depending on the maximum admitable error $\delta$. One can interpret the error as the closest rank circuit with removed T-gates from the initial circuit~\cite{Bravyi_2019}. In the case of low T-gates count, the circuit can still be efficiently simulated~\cite{Bravyi_2016}.

\paragraph{Clifford Point approximation} relies on projection from any kind of rotations on the Block sphere on the Clifford group.
\label{par:clifford_point}

\begin{figure}[H]
        \label{fig:ring_rotation}
        \centering
        \includegraphics[scale=0.15]{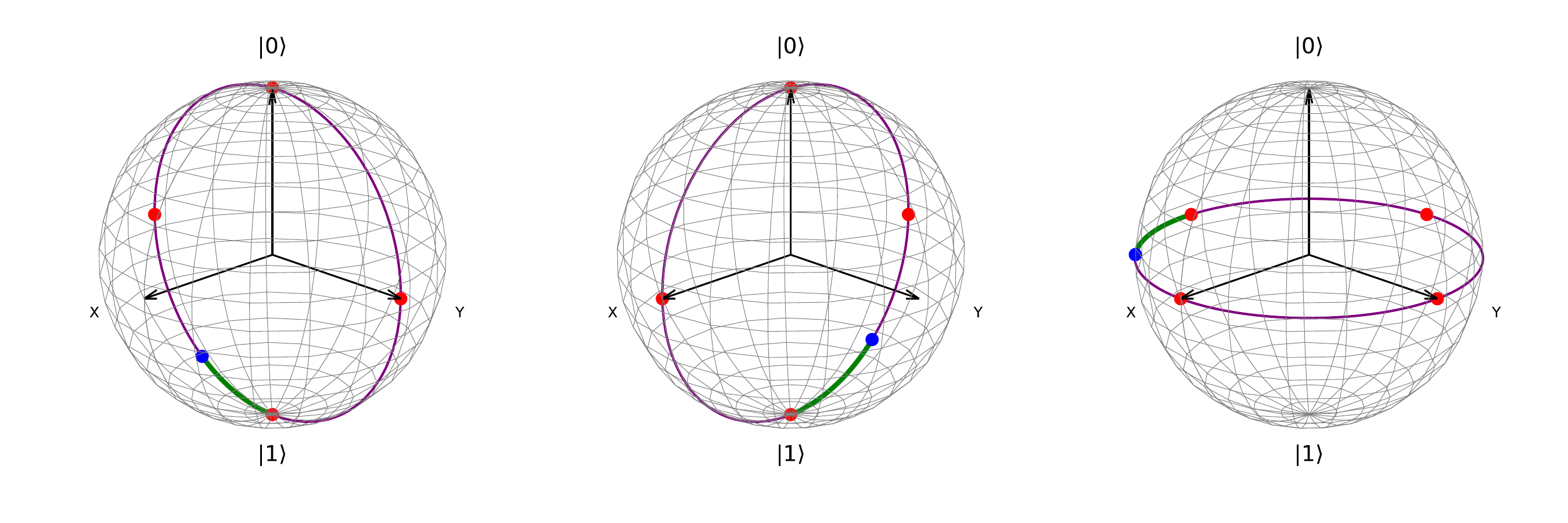}
        \caption{Bloch sphere representation of full Clifford space projection on each single qubit rotation}
\end{figure}

\begin{table}[ht]
    \centering
    \begin{tabular}{|c|c|c|}
        \hline
        Rotation & Angle & Clifford Projection \\
        \hline
        & $0$ & $I$ \\
        $\hat{R}_{x}(\theta)$& $\frac{\pi}{2}$ & $h$ \\
        & $\pi$ & $X$ \\
        & $\frac{3\pi}{2}$ & $h \, X$ \\
        \hline
        & $0$ & $I$ \\
        $\hat{R}_{y}(\theta)$& $\frac{\pi}{2}$ & $h$ \\
        & $\pi$ & $Y$ \\
        & $\frac{3\pi}{2}$ & $h \, Y$ \\
        \hline
        & $0$ & $I$ \\
        $\hat{R}_{z}(\theta)$& $\frac{\pi}{2}$ & $h$ \\
        & $\pi$ & $Z$ \\
        & $\frac{3\pi}{2}$ & $h \, Z$ \\
        \hline
    \end{tabular}
    \caption{Conversion Table from any rotation to approximate closest Clifford group gates. $h$ Hadamard gates, $(X, Y, Z)$ gates, and $I$ identity. Multi-qubit parametric gate are split into $CNOT$ single qubit rotations}
    \label{tab:conversion_table}
\end{table}
For any Pauly rotations $\hat{R}(\theta)$, we reduce the acceptable angles to the finite set $\{\theta\}_i\in[0, \pi/2, \pi, 3\pi/2]$, where $i$ denotes the parameter indices in the antsatz parameter vector. For any parametrized circuit containing only one-qubit rotations and CNOTs of the form Eq.~\eqref{eq:general_rotation}, each rotation is clipped to nearest members of the four elements set. An ADAPT antsatz being only made out of Pauli rotations and CNOTs, we can establish a lookup table of the form of Tab.~\ref{tab:conversion_table} for fast circuit transpilation. In the Bloch sphere representation, it only means that any rotation present on each of the $(\hat{R}_X(\theta), \hat{R}_Y(\theta), \hat{R}_Z(\theta))$ rings is projected on the contact points between the unit sphere and the reference axis Fig.~\ref{fig:ring_rotation}.

Let $\mathbf{x} = (x_1, x_2, \ldots, x_n)$ be a vector in $\mathbb{R}^n$. We are interested in the case where each element $x_i$ can take on one of $k$ discrete values from a set $S$. Define the set $S = { s_1, s_2, \ldots, s_k }$, and the subspace of vectors $\mathbf{x}$ can be written $\mathbf{x} \in \mathbb{R}^n \quad \text{such that} \quad x_i \in S \quad \text{for } i = 1, 2, \ldots, n$.

This leads to a total of $k^n$ possible combinations of the vector components. Thus, the total number of possible vectors $\mathbf{x}$ in this subspace is $N = k^n$. In summary, the subspace of vectors in $\mathbb{R}^n$ where each vector element can take on one of $k$ discrete values is defined as:

\begin{align*}
    \mathbf{x} \in (x_1, x_2, \ldots, x_n) \in \mathbb{R}^n &\mid x_i \in S \quad, i = 1, 2, \ldots, n, \\
    &\text{with} \quad N = k^n \nonumber
\end{align*}

We note that authors in~\cite{chengphysreview} show that for warm-starting, the initial point of the search space needs to be in the basis state. Along those Clifford Points, we can explore various discrete optimization methods. We introduce simple Hill-Climb (descent) algorithm with local refinement. The fine-tuning is tested with various discrete optimization methods in Tab~\ref{tab:discrete_methods}. We define the neighbor points as one step variation from one randomly picked variable direction, forming a simple local-search heuristics in parameter space. The fine-tuning allows to add variety in optimization methods, and allows to observe relationships between specific methods and exploration-exploitation tradeoff in the case of ADAPT-QAOA.

\begin{table}
    \centering
    \caption{Optimization methods explored}
    \label{tab:discrete_methods}
    \begin{tabular}{|p{3.5cm}|p{3.5cm}|}
        \hline
        Method                      & Description                                      \\ \hline
        Simulated Annealing~\cite{koch_sim}         & Mimics the annealing process to explore solution space.  \\ \hline
        Heuristic Random Local Search~\cite{koch_sim}          & Divides complex problems into simpler sub-problems. \\ \hline
        Ant Colony Optimization~\cite{ant_colony_parsons}        & Strategy inspired by ant foraging behavior, building solutions incrementally based on collective paths. \\ \hline
        Differential Evolution~\cite{koch_sim}      & Evolves candidate solutions using differences among them. \\ \hline
    \end{tabular}
\end{table}

\begin{figure*}
    \centering
    \vspace{-4cm}
    \includegraphics[scale=0.81]{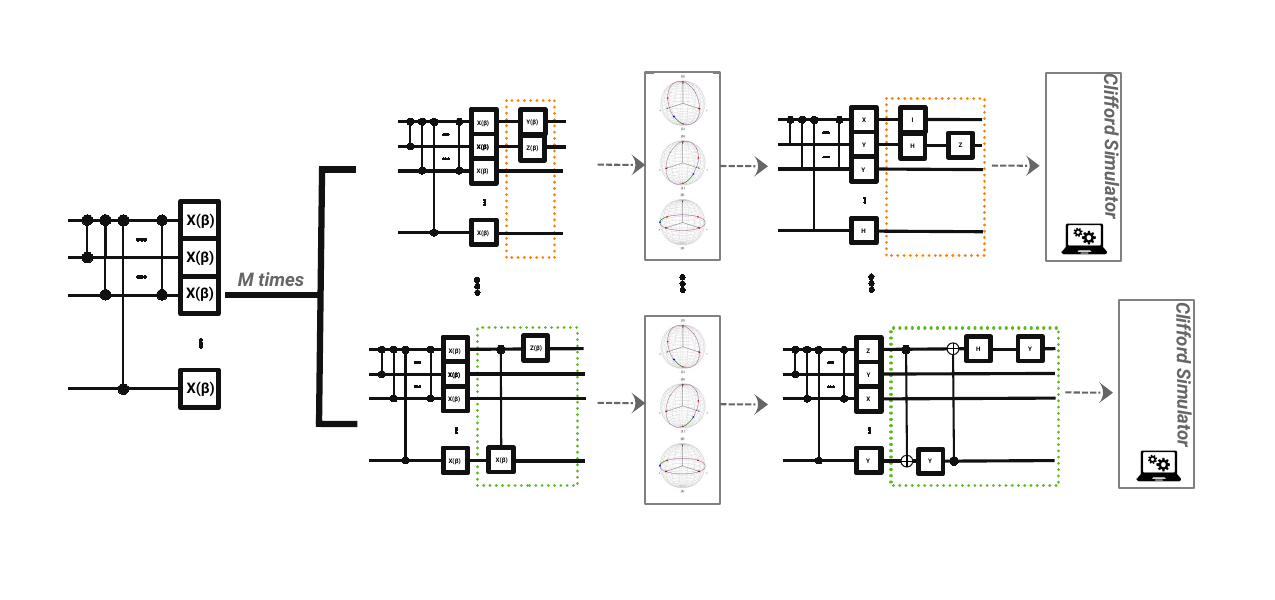}
    \caption{Composing Eq.~\eqref{eq:decision_problem_eval} using Clifford Point approximations allows for composing a fully parallel scheme in evaluating the direction of bigger change from the operator pool. Each elements from the operator pool of size $M$ can be simultaneously evaluated, expecting the classical resources to be readily available.}
    \label{fig:clifford_point_selection_parallel}
\end{figure*}

\paragraph{Low-rank stabilizer decomposition} relies on a suite of methods introduced in~\cite{Bravyi_2019}, composed in general in two steps: first the elaboration of exact or approximated low-rank stabilizer decomposition of the input circuit, then the application of efficient simulation algorithms. The method is articulated around the idea of \textit{approximate stabilizer rank}, the definition 2 of~\cite{Bravyi_2019} states: \textit{Suppose $\psi$} is a pure n-qubits state such that $||\psi|| = 1$. Let $\delta > 0$ be a precision parameter. The approximate stabilizer rank $\xi_\delta(\psi)$ is the smallest integer $k$ such that $||\psi -\psi'||\leq\delta$ for state $\psi'$ with exact stabilizer rank $k$. From this definition, authors develop a subroutine to manipulate such decompositions and define the tasks of finding on a quantum circuit $\hat{U}$, the classical simulation task of sampling bitstrings $x\in \{0, 1\}^n$ from $P_U(x) = |\langle x|\hat{U}|0^n\rangle|^2$. If given a subroutine providing an approximate stabilizer decomposition on the state $\hat{U}|0^n\rangle$:

\begin{equation}
    \label{eq:stabilizer_decomposition}
    ||\hat{U}|0^n\rangle - |\psi\rangle|| \leq \delta, \quad \psi\rangle = \sum_{\alpha=0}^kb_\alpha \hat{U}_\alpha|0^n\rangle
\end{equation}

Those procedures have been implemented in \textit{qiskit} with the Van den Nest Monte Carlo simulation~\cite{nest2010simulatingquantumcomputersprobabilistic}, providing multiple metropolis options, Markov chain time length, and a norm estimation method. Within the scope of this work, $\delta$ will be used as a control parameter as acceptable fidelity on the decomposition provided by the $\epsilon$-error parameter on the \textit{extended\_stabilizer} simulator.

\paragraph{Introducing classical approximations in ADAPT} can be done along multiple entry-points which we expose in Fig.~\ref{fig:flow_diagram_adapt_full}. In the following section we explore each of them through numerical evaluation. In \textcolor{green}{(1) Initial Optimization}, we initialize the QAOA instance from which the routine starts, we look into Clifford Point preoptimization, followed by either a continuous optimization or the low-rank decomposition approximation. In \textcolor{blue}{(2) Pool Operator Selection}, we investigate the role of errors in evaluating the expression Eq.~\ref{eq:decision_problem_eval}, applying directly Par.\ref{par:clifford_point} to nearest Clifford circuit as evaluation of the energy variation in the operator pool selection. This procedure is roughly presented in Fig.~\ref{fig:clifford_point_selection_parallel}, where we propose a parallel and fully classical method for selecting the ADAPT pool operator. Finally in \textcolor{orange}{(3) Post-operator Optimization}, we investigate the role of T-gates precision in solution quality at all levels of the objective evaluation.

In the results section Sec.~\ref{sec:results}, we look at the effects of Clifford Point preoptimization in the case of QAOA for both the MaxCut and the TFIM. We carry on by exposing the effects of selecting the best-fitting operator from the pool as a Clifford Point approximation. Finally, we expose experiments concerning the introduction of low-rank stabilizer decomposition on each circuit evaluation within the methods.

\section{Results}
\label{sec:results}
\subsection{Effects on the choice of discrete optimization strategy}
\label{secsec:choices_optimization_strategy}
In this section, we observe the effect of fined-tuned hill-climb Clifford Point preoptimization. Due to the theoretical and design properties of ADAPT, pre-optimizing -at least as a naive sequential approach- can only work at the ADAPT first iteration. This means that we explore the potential of preoptimization using Clifford Points, at setting up good initial conditions for ADAPT to improve systematically its convergence properties.

\begin{figure}[h]
    \centering
    \includegraphics[width=0.95\linewidth]{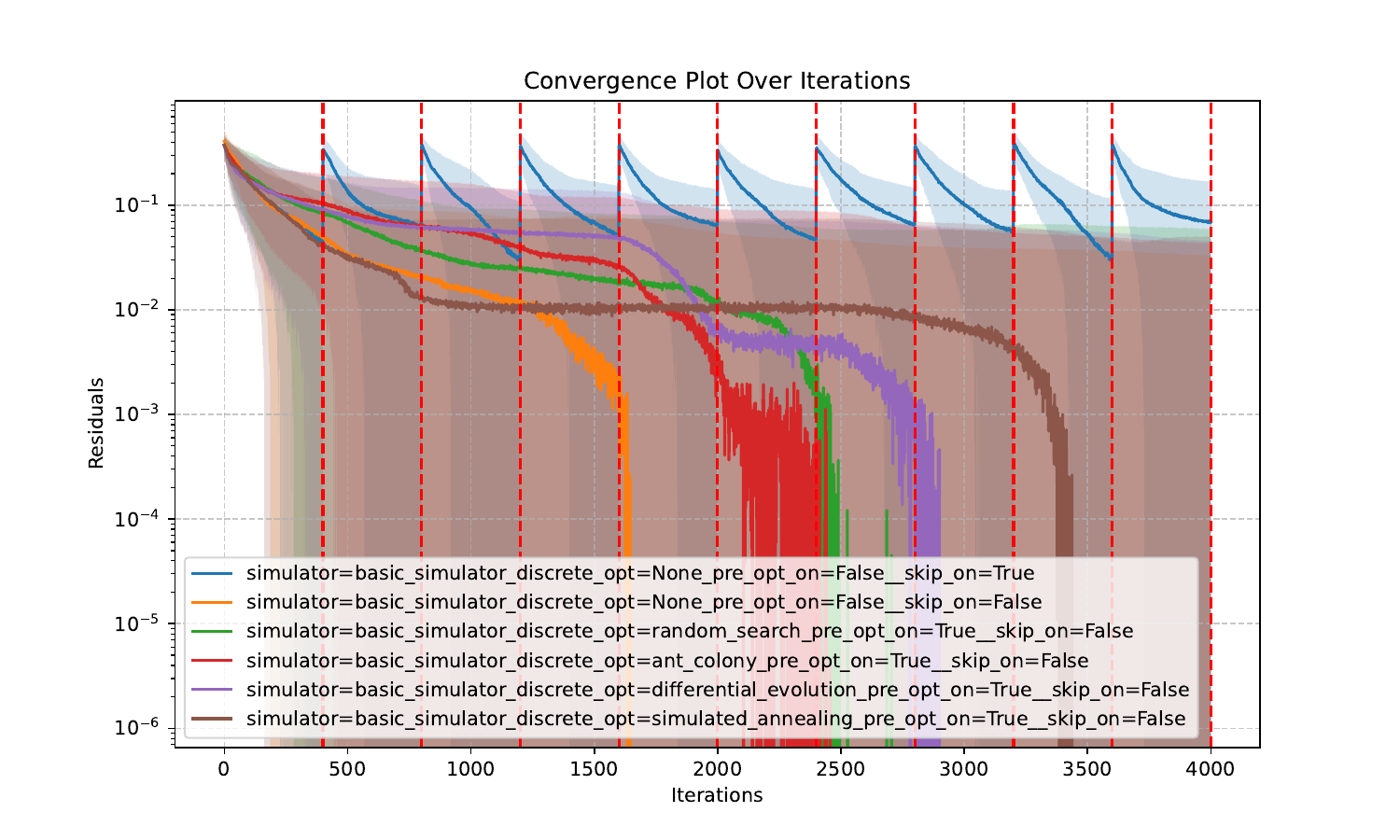}
    \caption{MaxCut, 30 runs with 10 ADAPT iterations. 300 SPSA iterations, reset point exploration for QAOA. No stopping conditions but the maximum iteration. 5 nodes random problems with fully connected graphs of random weights. Each pre-optimizer runs for 10 iterations with 5 fine-tuning hill climb iterations. ADAPT is fully simulated using \textit{basic\_simulator} in Qiskit.}
    \label{fig:maxcut_various_preopt}
\end{figure}

\begin{figure}[h]
    \centering
    \label{fig:clearer_picture_local_minima}
    \includegraphics[width=0.95\linewidth]{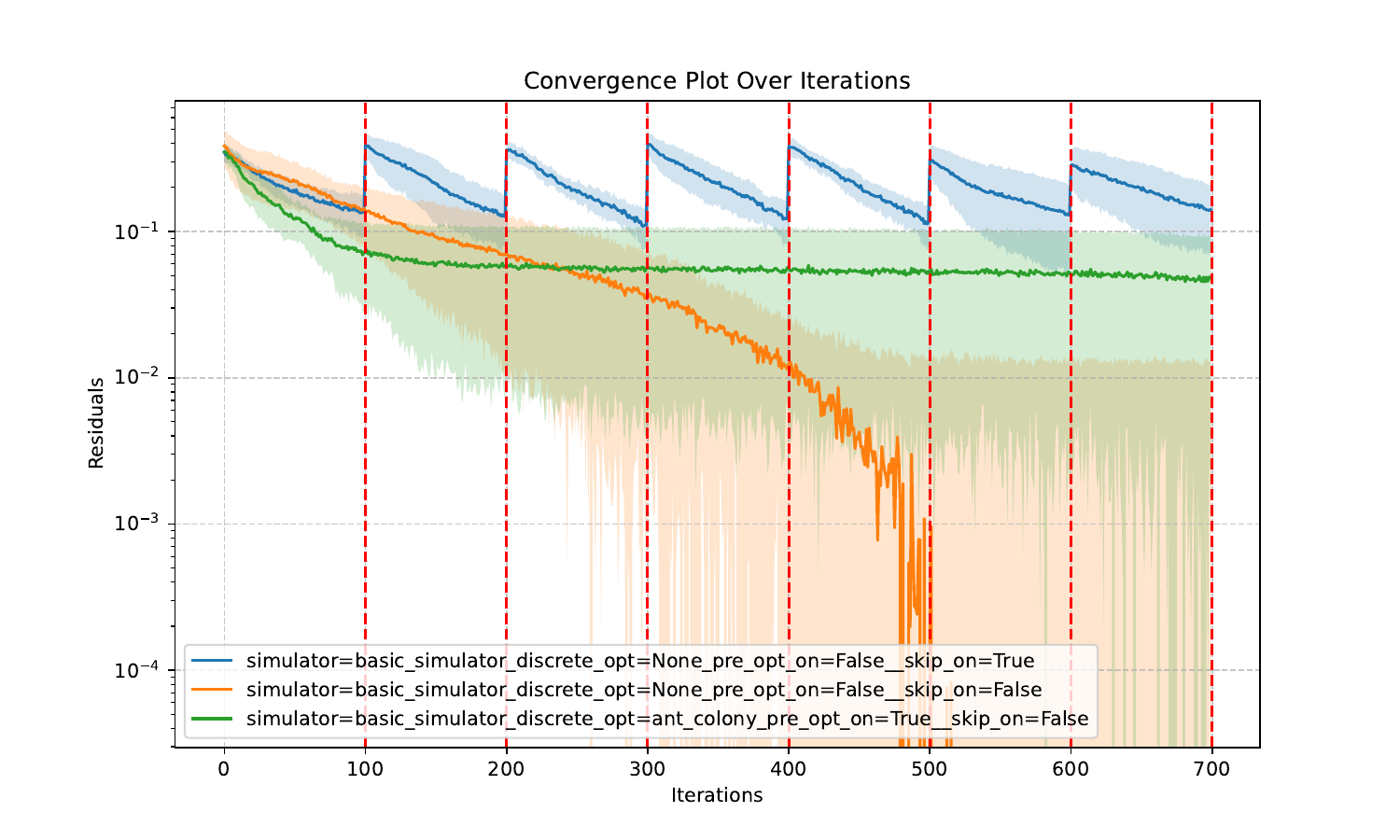}
    \caption{MaxCut, 30 runs with 6 ADAPT iterations. 300 SPSA iterations, reset point exploration for QAOA. Stopping condition on error at the floating point precision. 5 nodes random problems with fully connected graphs of random weights. Each pre-optimizer runs for 10 iterations with 5 fine-tuning hill climb iterations, on the ant colony pre-optimizer as example. ADAPT is fully simulated using \textit{basic\_simulator} in Qiskit.}
\end{figure}

In Fig.~\ref{fig:maxcut_various_preopt}, we show a naive result from 5 nodes random MaxCut problems, where all pre-optimizers are run with the same initial conditions as well as amount of iterations. Across multiple experiments, the solver shows high sensitivity to pre-optimization and initial conditions, showing oscillatory patterns, and very distinct improvement patterns. In this case, preoptimizing is detrimental to ADAPT. Even though Clifford Point preoptimizing can significative improve on initial convergence speed, it can drive ADAPT into a local minima, as shown more clearly in Fig.~\ref{fig:clearer_picture_local_minima}. This indicates that preoptimization strategies for MaxCut and ADAPT should integrate further information between the Clifford Point update and the continuous update (for example, by deploying momentum transfer strategies, or gathering information about the information space of the objective function during the Clifford Point optimization).

\begin{figure}[h]
    \centering
    \includegraphics[width=0.93\linewidth]{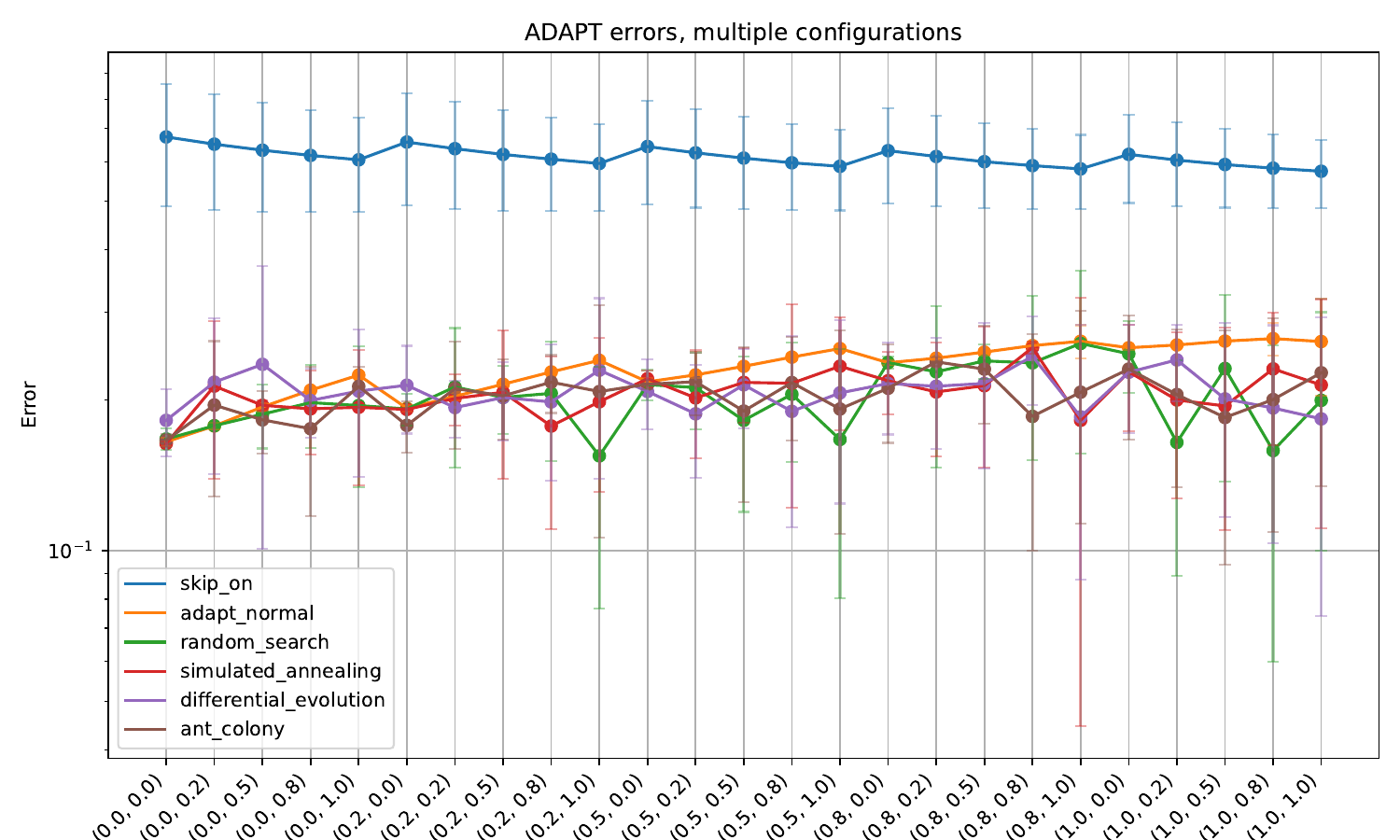}
    \caption{100 SPSA iterations, 10 Hill-Climb + 5 fine-tuning sweep Clifford Point optimization. 20 experiment reset, 5x5 cut of the $(g_x, g_z)$. Heterogeneity in initial conditions for ADAPT. The \textit{skip\_on} represents re-initialized QAOA, with no parametric gate update.}
    \label{fig:tfim_figure_error_section1}
\end{figure}

However, the TFIM problem shows more promising results in Fig.~\ref{fig:tfim_figure_error_section1}, where we observe that pre-optimization can systematically help with precision depending on (i) the chosen discrete optimization method, and (ii) the relative strength of the coupling parameters $(g_x, g_z)$. In particular, we observe an increased drift in preoptimizing and the $g_z$ coupling strength, where when $g_z\xrightarrow[]{} 0.0$, preoptimization provides the least improvements. An early explanation of this behavior can be found by looking at Eq.~\ref{eq:general_tfim_hamiltonian}, the larger the parameter, the largest are the contributions of the Hamiltonian into single qubits parametric-$Z_i$ gates. Due to the Clifford-Point projection on the Z-basis, T-gate losses errors around $Z$-rotations results in lowered overall errors made in the continuous optimization, as the discrete Clifford space explored maximizes the grid matching when $Z_i$ gates are present predominately in the circuit (or their contributions are made more important in the shape of the objective function). When only two-qubits gates are presents at $(0.0, 0.0)$, preoptimization is counterproductive, when it increases in positive effect with the contributions of single-qubit gates.

MaxCut being a subset of the TFIM problem for the specific $(g_x, g_z) = (0.0, 0.0)$, it seems than ADAPT with Clifford Point approximation would benefit from additional operator selection heuristics, favoring single qubits operators at antsatz design. Regardless, in the following section, we look into how one can introduce those approximations within the operator selection step of the ADAPT procedure.

\subsection{Clifford-only ADAPT operator selection}
\label{secsec:clifford_selection_operator}

During the ADAPT procedure, in Fig.~\ref{fig:flow_diagram_adapt_full} we write $\Delta E$ the vector elements associated to each operator indices, described formally in Eq.~\ref{eq:decision_problem_eval}. For each element of the operator pool, the task is to estimate the commutator between the previous iteration antsatz, and every possible operator, then picking the direction generating the greater change. We introduce a Clifford Point approximation at each evaluation of the commutator, bringing the amount of quantum circuits calls on hardware to zero in $\Delta E$ evaluation. We explore numerically the step \textcolor{blue}{2} in Fig.~\ref{fig:flow_diagram_adapt_full}, for the MaxCut and TFIM problems.

\begin{figure}[h]
    \centering
    \includegraphics[width=0.93\linewidth]{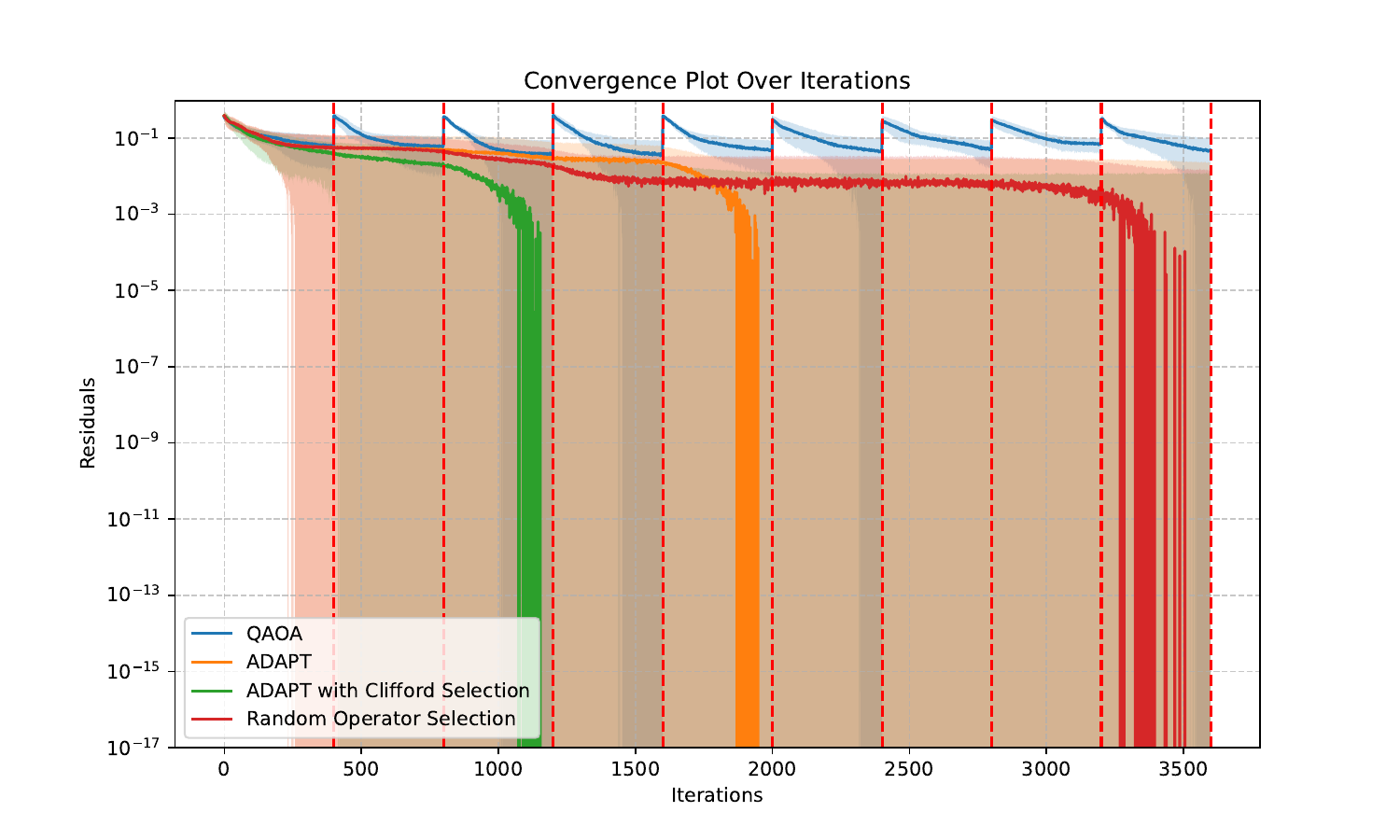}
    \caption{MaxCut, Clifford Accelerated ADAPT, random operator selection. 6 Nodes, 10 ADAPT iteration, 400 SPSA iterations.}
    \label{fig:demo_selection_6_nodes}
\end{figure}

Results in Fig.~\ref{fig:demo_selection_6_nodes} show convergence behavior at much lower parameter count when Clifford Point Selection is used. This is a strong indication that for the MaxCut Hamiltonian structure, the Clifford approximation makes better decisions in extending the QAOA mixer layer. A possible explanation relies on a similar reason than preoptimization was failing for MaxCut: local errors in the operator selection combined to the symmetries in MaxCut improves on global operator choice, while averaging out local information due to the projection onto Clifford Points.

\begin{figure}[h]
    \centering
    \includegraphics[width=.9\linewidth]{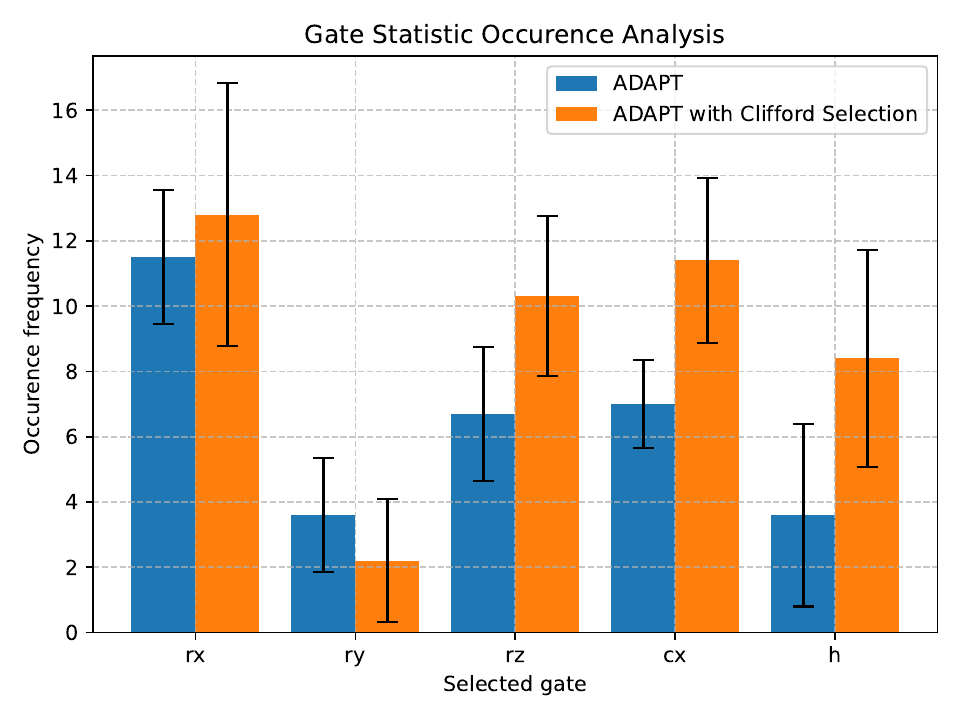}
    \includegraphics[width=.9\linewidth]{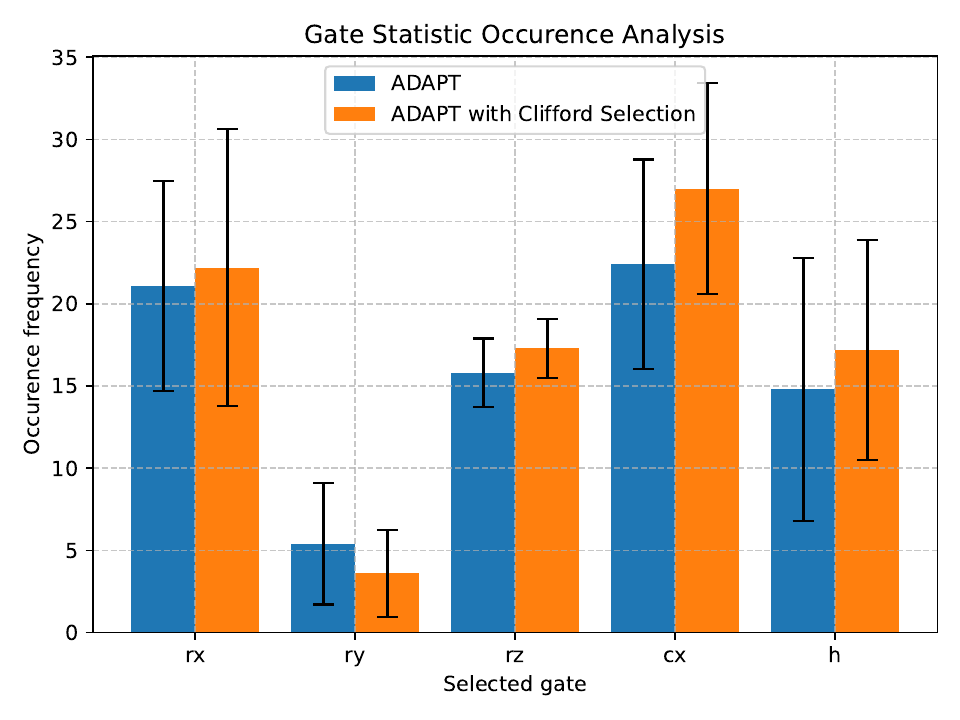}
    \caption{50 randomized runs at 400 iterations SPSA, 15 ADAPT runs, result on added gate for random MaxCut problems (3 nodes top, 6 nodes bottom)}
    \label{fig:maxcut_gate_selection}
\end{figure}

Each ADAPT iteration introduces a new parametric gate and the ADAPT procedure tends to choose two qubit gates over single qubit gates. In Fig.~\ref{fig:maxcut_gate_selection}, a simple count of the selected gates over multiple problems and optimization show an increase in $RRZ$ gates, to the detriment of $RY$ rotations when using the Clifford Point selection. This indicates that the Clifford Approximation is compensated by the introduction of more controlled-Z rotations in the mixer layer of the antsatz, which is overall a better decision for expressivity of the circuit in accord with~\cite{zhu2022adaptivequantumapproximateoptimization}.

\begin{figure}[]
    \centering
    \includegraphics[width=0.9\linewidth]{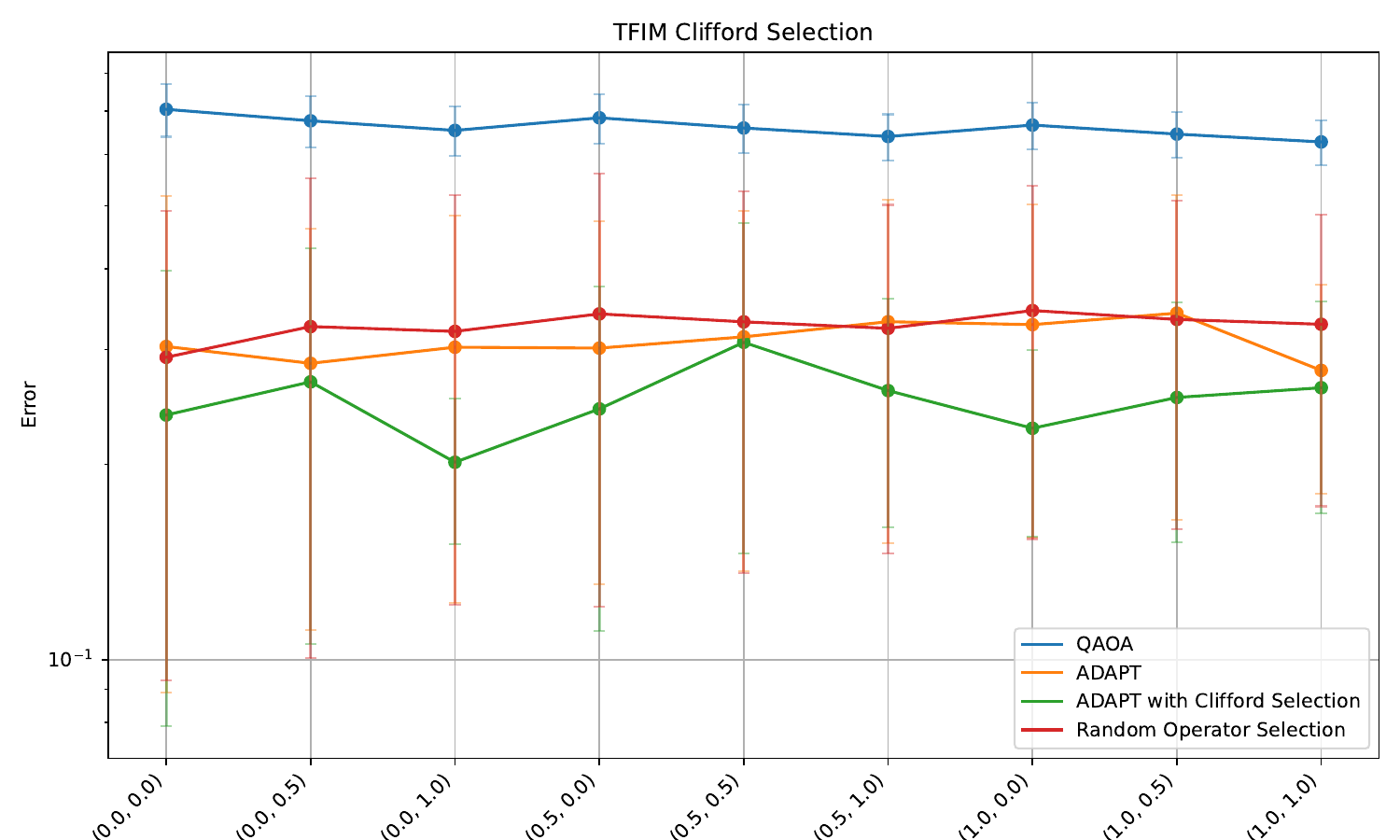}
    \caption{ADAPT, Clifford Accelerated ADAPT, random operator selection. 5 qubits Ising chain, 10 ADAPT iteration, 400 SPSA iterations for the TFIM problem. 3x3 discretization of the control-couple $(g_x, g_z)$)}
    \label{fig:tfim_clifford_selection}
\end{figure}

The TFIM problem tells a similar story in Fig.~\ref{fig:tfim_clifford_selection}, where the Clifford selection outperforms other operator selection. We note on this picture that the random operator selection appears better in some places, but the variance in behavior is more than twice larger than ADAPT. This behavior is not repeated for larger problems, but where it becomes much more expensive and time consuming to gather significant amount of data.

\subsection{Low-rank stabilizer decomposition}
\label{secsec:low_rank_decomposition_proof}
This last section observes the behavior of ADAPT to the same previous problems while introducing low-rank decomposition approximations at objective function evaluation. In particular we look at how much error on T-gates (hence overall in our case, on $Z$-rotations) can be tolerated without a significative loss in performances. Providing the fidelity introduced in Eq.~\ref{eq:stabilizer_decomposition} as a control variable, we observe capabilities of ADAPT in finding solutions for MaxCut and the TFIM problem.

\begin{figure}
    \centering
    \includegraphics[width=0.9\linewidth]{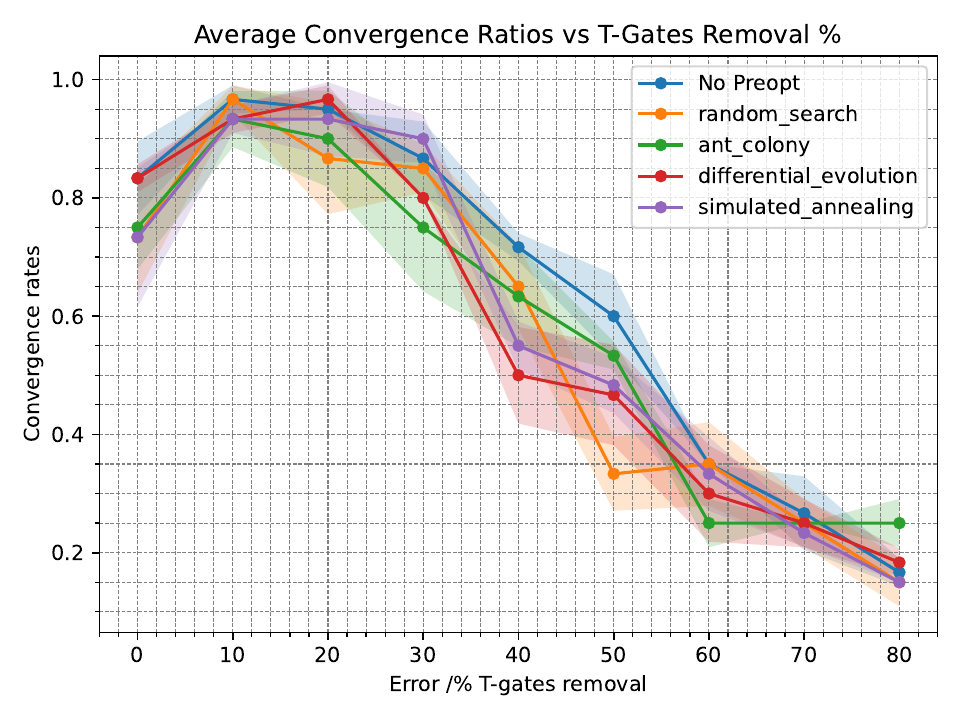}
    \caption{Random fully connected 5 nodes graphs 20 runs on 20 problems, 300 iterations of SPSA using standard hyper-parameters definitions. 10 iterations of fined-tuned hill-climb, 5 iterations fine-tuning. 1\% tolerance as a target.}
    \label{fig:t_gates_diminish_rates}
\end{figure}

\begin{figure}
    \centering
    \includegraphics[width=0.9\linewidth]{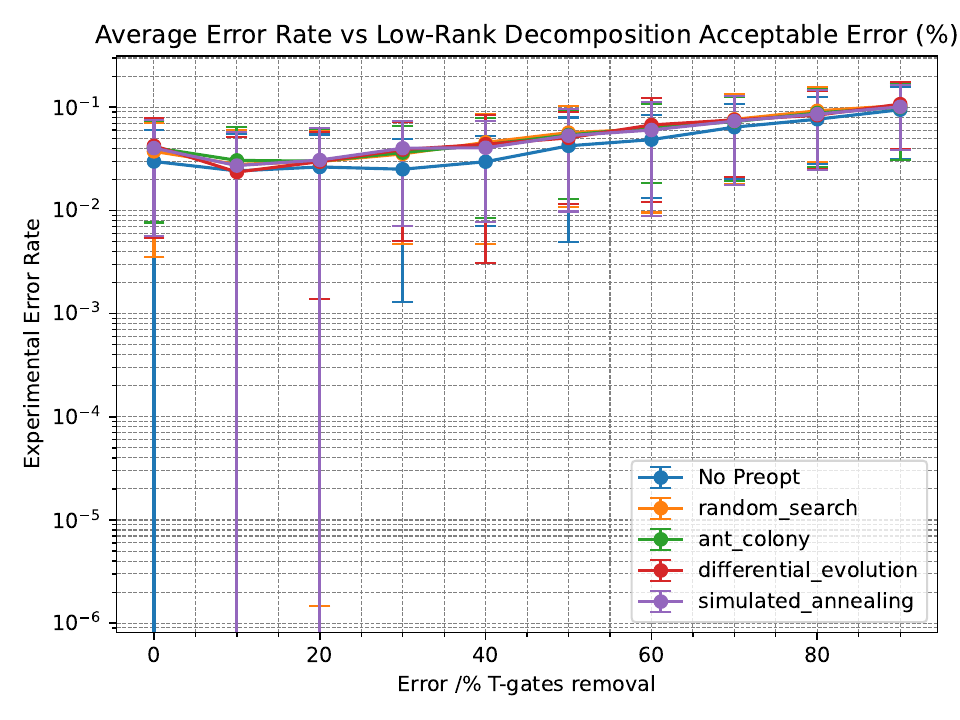}
    \caption{Random fully connected 5 nodes graphs 20 runs on 20 problems, 300 iterations of SPSA using standard hyper-parameters definitions. 10 iterations of fined-tuned hill-climb, 5 iterations fine-tuning. 1\% tolerance as a target.}
    \label{fig:t_gates_diminish_error}
\end{figure}

Our results in Fig.~\ref{fig:t_gates_diminish_rates} shows that similarely to Sec.~\ref{secsec:choices_optimization_strategy}, pre-optimization does not help much on MaxCut regarding convergence behaviors. This means that Clifford Point preoptimization seems to not be effective at offsetting T-gate fidelity errors. However, we observe that optimization improves when 10\% to 30\% precision on T-gates is lost. This is validated further on Fig.~\ref{fig:t_gates_diminish_error}, looking at relative errors to ground truth, rather than 1\% tolerance stopping conditions, where action of the pre-optimizers are as well providing mixed results.

\begin{figure}
    \centering
    \includegraphics[width=0.9\linewidth]{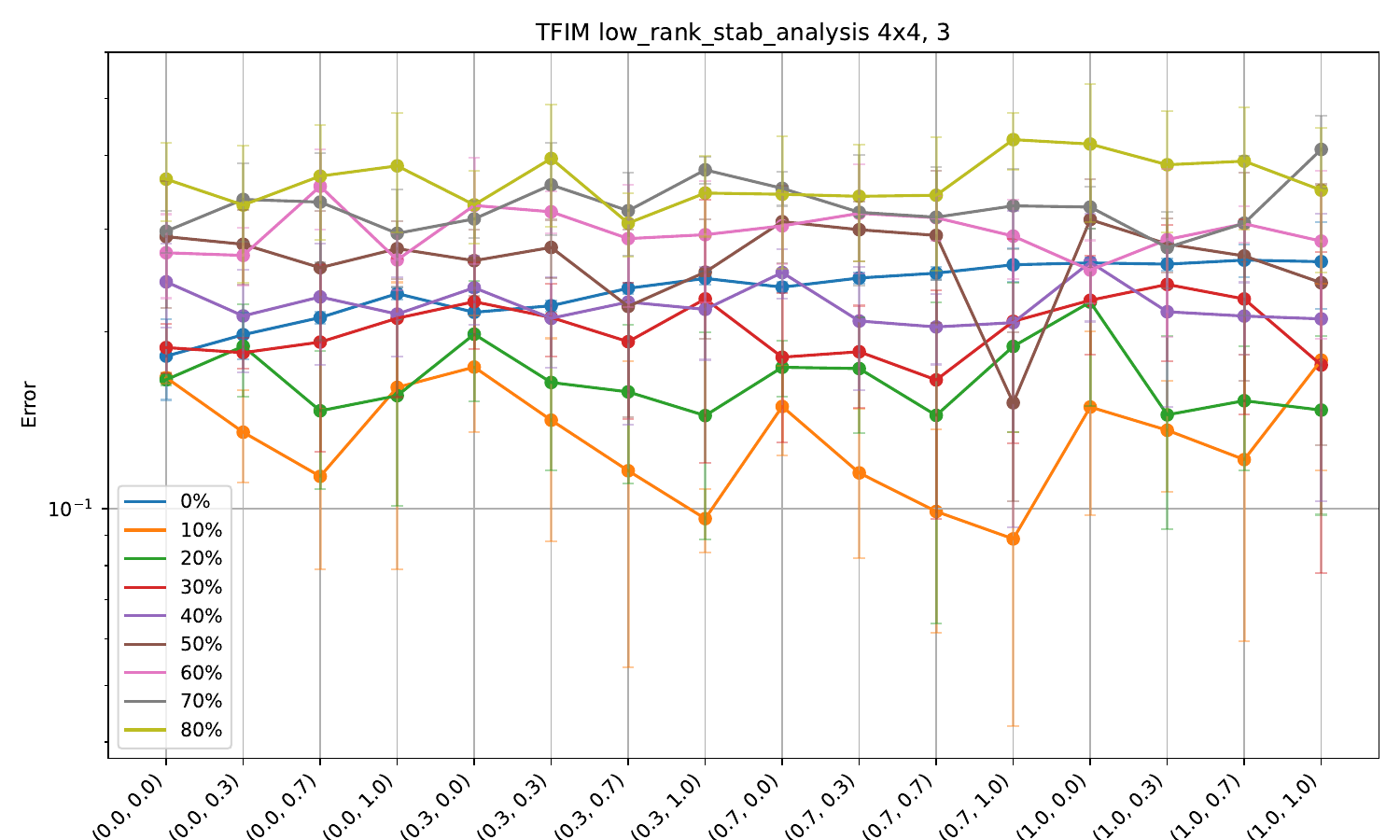}
    \caption{TFIM solver on 5 qubits Ising Chain, 200 SPSA iterations using \textit{extended\_stabilizer}, 50 experiments ran on 256G of RAM, 15ms Monte Carlo mixing time. Percentages represent the acceptable fidelity on the low-rank decomposition. 4x4 discretization of the control-couple $(g_x, g_z)$)}
    \label{fig:tfim_3_nodes4_4_stabilizer}
\end{figure}

This improvement, is even more striking when looking at the TFIM problem in Fig.~\ref{fig:tfim_3_nodes4_4_stabilizer}, where losing 10 to 20\% of T-gate precision largely improves performances. This striking observation can be explained by two ways: (l) the initial antsatz is over-represented in T-gates, or (2) the classical continuous search method is biased towards geometrical properties of the objective function around non-Clifford point values. For all experiments, we note that the amount of shots is 512 regardless of the method used.

\section{Conclusion}
\label{sec:conclusion}
In this paper, we approached the ADAPT-QAOA solver using \textit{statevector} and \textit{extended stabilizers} simulator to evaluate the potential of introducing Clifford Point and low-rank decomposition approximations in the ADAPT-QAOA solver of MaxCut and the TFIM. In this context, we introduced three entry points, first the preoptimization of the QAOA routine, which showed counter-productive for the MaxCut problem, but useful for the TFIM as the control-parameter $g_z$ increases. Second we introduced the Clifford Point operator selection, which shows great promise for both the MaxCut and the TFIM, we showed that the approximation subtlety changes the ADAPT behavior, selecting on average more $RZZ$ gates, to the detriment of single qubit $RY$ gates. This method improves the operator selection procedure, while bringing QPU calls to zero, as every Clifford circuit evaluation can efficiently be sampled on classical hardware. In addition, the method is fully parallelizable. This approach opens the door for further quantum-classical integration, where every operation not providing quantum speedup should be run on classical hardware. Finally, we showed that using the \textit{extended\_stabilizer} using the fidelity on T-gates as a control parameter, we observed that convergence was significantly improved around 10\% to 30\% precision losses in T-gates evaluations in the low-rank stabilizer decomposition.

The latter results could be exploited in antsatz design on one hand or classical optimizer on the other hand: hand-crafted methods leveraging a subspace limiting T-gates operations, or the initial antsatz design is sub-optimal in T-gates rotations, where further compilation techniques could be introduced artificially removing unnecessary non-Clifford operators with performance improvements.

Further investigation should be done on multiple levels, first the same base approach could be implemented on but this approach could be applied on other reconfigurable variational quantum algorithms, such as DC-QAOA~\cite{Hegade_2022}, or ADAPT-VQA~\cite{Grimsley_2019}. The latter, especially on chemistry problem, is likely to show the most advantage with the Clifford Point approximations, motivated by the results for the classic VQA in~\cite{chengphysreview}. Further scaling investigation should be investigated, as to uncover if those improvements remain at larger scales.

Further theoretical investigation is necessary in finding theoretical limits and rigorious explanations around the multiple approximations. In particular the Clifford Point operator selection: it is likely that the properties observed are direct consequences of the Hamiltonian form Eq.~\ref{eq:general_tfim_hamiltonian} (MaxCut being a special case of this form), but might not show in chemistry problems. If possible, an explicit link between Clifford Points and those expressions using Eq.\ref{eq:decision_problem_eval}, and Eq.~\ref{eq:general_tfim_hamiltonian} should be also the subject of further research.

\section*{Acknowledgments}
This research was made possible by the financing of the BMWK-Project »EniQmA« for the Systematic Development of Hybrid Quantum Computing Applications.

\appendix
\onecolumn
\end{document}